\documentclass[prd,twocolumn,showpacs,preprintnumbers,floats,floatfix,nofootinbib]{revtex4}

\usepackage{graphicx}
\usepackage{dcolumn}

\begin{document}

\title{Forming Nonsingular Black Holes from Dust Collapse}
\author{R. Maier and I. Dami\~ao Soares}
\hspace{0.5cm}

\affiliation{Centro Brasileiro de Pesquisas F\'\i sicas, Rua
Dr. Xavier Sigaud 150, Urca
\\Rio de Janeiro. CEP 22290-180-RJ, Brazil}

\date{\today}

\begin{abstract}
The dynamics of the gravitational collapse is examined in the realm
of string based formalism of D-branes that encompass General Relativity as a low energy limit.
A complete analytical solution is given to the spherically symmetric collapse of a pure dust star,
including its matching with a corrected Schwarzschild exterior spacetime.
The collapse forms a black hole (an exterior event horizon)
enclosing not a singularity but perpetually bouncing matter in the infinite chain
of spacetime maximal analytical extensions inside the outer event horizon.
This chain of analytical extensions has a structure analogous to that of the Reissner-Nordstrom
solution, except that the timelike singularities are avoided by bouncing barriers.
The interior trapped bouncing matter has the possibility of being
expelled by disruptive nonlinear resonance mechanisms.
\end{abstract}

\label{PACS number:}
\maketitle
\newpage
\par
{\it{Introduction}} -
Black holes are solutions of vacuum General Relativity equations describing
the exterior spacetime of the final stage of gravitationally bounded systems
whose masses exceeded the limits for a finite equilibrium configuration\cite{chandra}.
Geometrically a black hole may be described as a region of asymptotically flat
spacetimes bounded by an event horizon hiding a singularity formed in the collapse.
Fundamental theorems by Israel and Carter\cite{carter} state that
the final stage of a general collapse of uncharged matter is typically a
Kerr black hole, which has an involved singularity structure.
Nevertheless, for a realistic gravitational collapse we have no evidence that the Kerr solution
describes accurately the interior geometry of the black hole. On the contrary, the best
theoretical evidence presently available indicates that the interior of the black hole
thus formed is analogous to the interior of a Schwarzschild black hole with a global
spacelike singularity\cite{wald}. The simplest way of forming such structure is by
the spherical collapse of dust, as originally shown in the classical paper of
Oppenheimer and Snyder\cite{oppen}. However, as singularities cannot be empirically conceived,
this turns out to be a huge pathology of the theory.

\par
Albeit the cosmic censorship conjecture\cite{wald1},
there is no doubt that the General Theory of Relativity must be properly corrected or
even replaced by a completely new theory, let us say a quantum theory of gravity.
This demand is in order to solve the issue of the presence of singularities predicted by
classical General Relativity, either in the formation of a black hole or in the
beginning of the universe. While a full quantum gravity theory remains presently
an elusive theoretical problem, quantum gravity corrections near singularities
have been the object of much recent research, from loop quantum cosmology\cite{bojo} to
string-based formalism of D-branes theory\cite{string}.
In the latter scenario extra dimensions are introduced constituting the bulk space.
All matter would be trapped on a 4-dim world-brane spacetime embedded in the bulk
and only gravitons would be allowed to move in the full bulk. At low energies General
Relativity is recovered\cite{rs} but at high energy scales
significant changes are introduced into the gravitational dynamics and the singularity can be removed.
\par In this context, we examine the gravitational collapse of a pure dust star in the realm
of braneworld theory in a 5-dim bulk, a procedure analogous to that of
Oppenheimer and Snyder\cite{oppen}. We reproduce the interior
Friedmann-Robertson-Walker solution
and obtain its analytical matching with a corrected Schwarzschild exterior geometry.
The maximal analytical extension of such geometry is analogous to that of the Reissner
Nordstrom solution but now with no charge and no singularity at all.
Our configuration exhibits at most two event horizons
enclosing perpetually bouncing matter with no singularity formation. We also examine the
implications of such corrections in the context of the General Relativity tests, and in the
Hawking evaporation of black holes. We rely basically on Refs. \cite{maeda} for the modified Einstein's
equations in a 4-dim braneworld embedded in a 5-dim bulk. Related literature on the problem
of nonsingular black holes is properly addressed at the end of the paper. Throughout the paper
we use units such that $G_{N}=c=1$; however for clarity we maintain $G_{N}$ in all expressions.
\par
{\it{The Model}} -
We assume a spacetime braneworld model embedded in a 5-dim bulk with a timelike extra dimension, whose
matter content is a spherically symmetric collapsing dust with density $\rho$. In a coordinate system comoving with dust the interior geometry is still shown to be a Friedmann-Robertson-Walker metric\cite{weinberg}
\begin{eqnarray}
\label{eq1}
d\tau^2=-g_{ab} dx^a dx^b=dt^2-a^2(t)\Big(\frac{1}{1-kr^2}dr^2-r^2d\Omega^2\Big)~~
\end{eqnarray}
with its dynamics given by the first order modified Friedmann equation
\begin{eqnarray}
\label{eq2}
\dot{a}^2=- k +\frac{8\pi G_{N} E_{0}}{3a}-\frac{4\pi G_{N} E^2_{0}}{3|\sigma| a^4}
\end{eqnarray}
where $E_0$ is a constant of motion associated with the dust density, $E_0=\rho a^3$, and $\sigma$ is the
negative brane tension\cite{comment}. $G_N$ is Newton's constant on the brane.
Assuming initial conditions for the collapse $\dot{a}(0)=0$
and $a(0)=1$, we get
\begin{eqnarray}
\label{eq3}
k=\frac{8 \pi G_N}{3}\Big[E_0-\frac{E_0^2}{2 |\sigma|} \Big].
\end{eqnarray}
The last term in the RHS of Equation (\ref{eq2}) is a correction due to the bulk-brane interaction
that eventually avoids the singularity, acting as a bouncing barrier. Its negative sign is
consequence of the timelike character of the extra dimension. The scale of the bounce is given by
$\rho \sim 2 |\sigma|$ that corresponds to a minimum of the scale factor $a_m \sim \Big(E_0/2|\sigma| \Big)^{1/3}$.
\par We determine the spherically symmetric metric outside the collapsing star from its
matching to the metric (\ref{eq1}), at the surface defined by $r=\gamma$ in comoving coordinates.
To this end let us transform the comoving coordinates of (\ref{eq1}) to Schwarzschild coordinates
($T,R,\bar{\theta},\bar{\phi}$) through the equations
\begin{eqnarray}
\label{eq4}
T=F[S(r,t)],~~R=a(t)~r,~~\bar{\theta}=\theta,~~\bar{\phi}=\phi,
\end{eqnarray}
with $F[S(r,t)]$ an arbitrary function of $S$, where $S(r,t)$ satisfies
\begin{eqnarray}
\label{eq5}
\Big[1-\frac{\dot{a}^2 R^2}{a^2-k R^2} \Big]\Big(\frac {\partial S}{\partial t}\Big)^{-1} \Big(\frac {\partial S}{\partial R}\Big)+\Big[-\frac{a\dot{a} R}{a^2-k R^2} \Big]=0~~
\end{eqnarray}
The line element (\ref{eq1}) is diagonal in this new coordinate system and we automatically
guarantee that
\begin{eqnarray}
\label{eq6}
\nonumber
g_{R R}&=&\Big[1-r^2(k+\dot{a}^2)\Big]^{-1}=\\
&&\Big[1-\frac{8 \pi G_N E_0 r^2}{3a}
+\frac{4 \pi G_N E_{0}^{2}r^2}{3|\sigma|a^4}\Big]^{-1},
\end{eqnarray}
where the second equality follows from Equation (\ref{eq2}). It is straightforward to verify
that the solutions for (\ref{eq5}) are
\begin{eqnarray}
\label{eq7}
S(r,t)=\delta+\alpha(1-kr^2)^{\frac{1}{2}}\exp{\Big(-k\int \frac{1}{a\dot{a}^2}da\Big)},~ k\neq0~
\end{eqnarray}
and
\begin{eqnarray}
\label{eq8}
S(r,t)=\delta+C\Big(\frac{r^2}{2}+ \int \frac{1}{a\dot{a}^2}da\Big),~~~~  k=0,
\end{eqnarray}
\begin{figure}
\begin{center}
\hspace{-0.5cm}
{\includegraphics*[height=7.8cm,width=8.9cm]{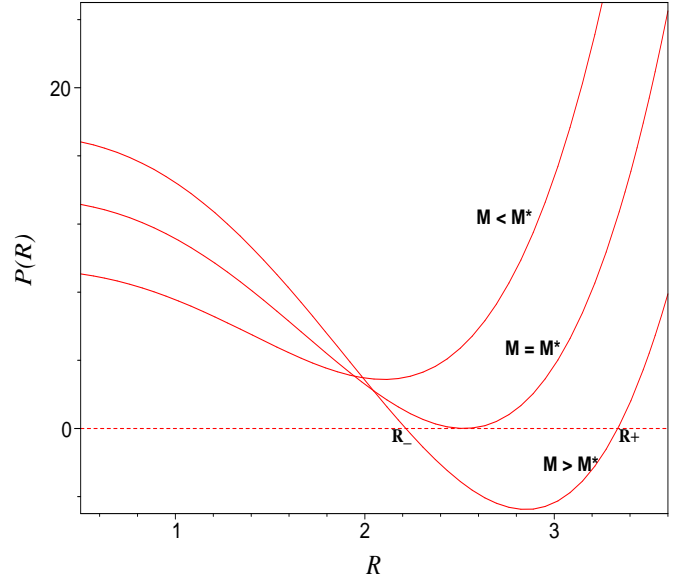}}
\vspace{0.2cm} \caption{Plot of the polynomial
${\mathcal P}(R) \equiv R^4 g_{T T}$ for dust masses $M<M_{\ast}$
(no black hole), $M=M_{\ast}$ (extremal black hole) and $M>M_{\ast}$ (black hole with outer horizon
${R}_{+}$ and inner horizon ${R}_{-}$). The figure corresponds to $|\sigma|=0.05$, in units $G_N =c=1$.}
\label{fig1}
\end{center}
\end{figure}
where $\delta$, $\alpha$ and $C$ are arbitrary constants. With the use of (\ref{eq2}) the above
integrals are analytically soluble for $k \geq 0$. For the physical domain of parameters
to be considered here, $S$ turns out a monotonous function of $a$ which can be properly inverted
to express $a$ in terms of $S$, for an explicit range of $a$. The remaining task is to choose
$F$ in terms of $S$. A choice can be suitably made for the case $k >0$
\cite{maier} so that at the
surface of matching $r=\gamma$ we obtain
\begin{eqnarray}\
\label{eq9}
\nonumber
d \tau^{2}&=&\Big(1-\frac{2G_N M}{R}
+\frac{3 G_N M^{2}}{4 \pi |\sigma|R^4}\Big)~d T^2\\ &-& \Big(1-\frac{2G_N M}{R}
+\frac{3 G_N M^{2}}{4 \pi |\sigma|R^4}\Big)^{-1}~d R^2 - R^{2} d {\Omega}^{2},~~~~~
\end{eqnarray}
where
\begin{eqnarray}\
\label{eq10}
M \equiv \frac{4 \pi}{3}~E_0 \gamma^3
\end{eqnarray}
is the total mass of the collapsing dust. The equation of the surface of the star is given
by $R=a(t)~\gamma$. In the remaining of the paper we are restricted to $k>0$. In this
instance the motion of dust is oscillatory, bouncing between $a=1$ and $a_m$ -- the two real
roots of $\dot{a}^2=0$ in Eq. (\ref{eq2}). The assumed initial conditions for the collapse,
$a(0)=1$ and $\dot{a}(0)=0$, require then that $a_m <1$, leading to a restriction on $\sigma$,
namely $|\sigma|> 2 E_0$.
\par
\begin{figure}
\begin{center}
\hspace{-0.5cm}
{\includegraphics*[height=9.9cm,width=8.0cm]{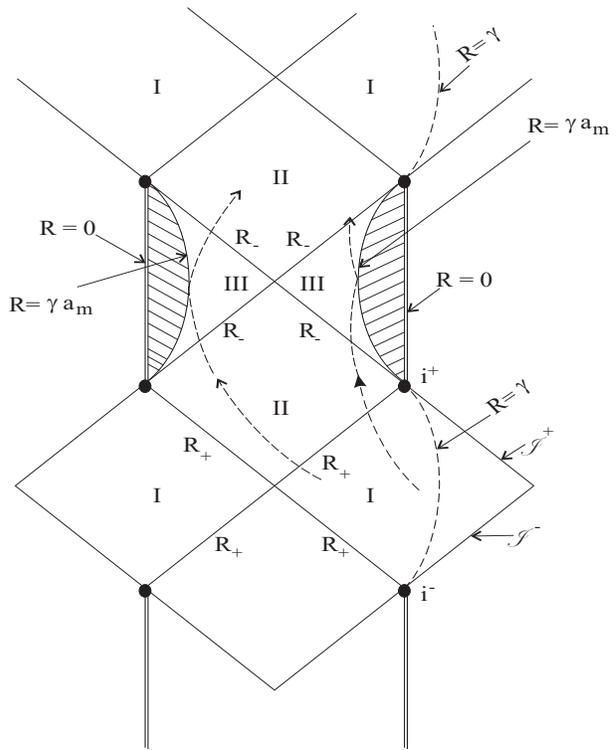}}
\vspace{0.2cm} \caption{Penrose diagram for the spacetime (\ref{eq9}) assuming $M > M_{\ast}$ (cf. (\ref{eq11})). The infinite chain of asymptotically flat regions I $(\infty>R>R_{+})$ are connected to regions III $(R_{-}>R>\gamma a_{m})$ by regions II $(R_{+}>R>R_{-})$. The dashed portion of regions III bounded by $R=\gamma a_{m}$
and $R=0$ is not part of the spacetime. Once crossing $R_{-}$ (the Cauchy horizon) the matter remains perpetually bouncing between $R=\gamma$ and $R=\gamma a_{m}$. Dashed arrowed lines illustrate timelike geodesics of dust particles and observers bouncing at $R=\gamma a_{m}$.}
\label{fig2}
\end{center}
\end{figure}

{\it{Maximal Extension}} -
In order to examine the maximal extension of the exterior geometry (\ref{eq9}), we need to
know whether, and under what circumstances, the configuration forms event horizons.
A necessary condition is that the mass $M$ of the collapsing star equals or exceeds
a critical limit $M_{\ast}$~,
\begin{eqnarray}
\label{eq11}
M \geq M_{\ast} \equiv \Big( \frac{4}{9 \pi G_N^3~|\sigma|}\Big)^{1/2}~,
\end{eqnarray}
($|\sigma|$ fixed), namely, that the polynomial ${\mathcal P}(R) \equiv R^4 g_{T T}$ has
one ${R}_{\ast}$, or two (${R}_{-} < {R}_{+}$) roots respectively (cf. Fig. 1).
Otherwise we cannot have formation of event horizons. The maximal analytical extension for the spacetime (9)
is considered for condition $M>M_{\ast}$ (cf. (11)) and $\gamma a_m < R_{-}$~. It is analogous
to that of a Reissner-Nordstrom black hole\cite{chandras} for $Q^2<M^2$ but without timelike singularities
(cf. Fig. \ref{fig2} and its caption). It is worth remarking that the surface of the collapsing
dust -- once crossing $R_{+}$ -- must necessarily cross ${R}_{-}$~, namely,
$\gamma a_{m}< {R}_{-}$~, so that a {\it stable} black hole
forms with trapped perpetually bouncing matter. We should mention that extensive computer tests
of the matching (\ref{eq4})-(\ref{eq9}) with parameters taken in a physically satisfactory domain
(basically in accordance with (\ref{eq11}), and satisfying the lower bound condition for $|\sigma|$
in order to have oscillatory motion) necessarily lead to $\gamma  a_{m}< R_{-}$, assuring that
Eq. (\ref{eq11}) is in fact a consistency condition for the formation of a nonsingular black hole
with two horizons. For illustration let us consider $\delta=1$ and $\alpha=0.01$ in (\ref{eq7}).
We also take $E_0=1.04 \times 10^{-21}~m^{-2}$ (corresponding to one thousand times the solar mass density),
$\gamma=6.96 \times 10^{8}~m$ (corresponding to one solar radius) and $\sigma=10^8$ (a large value
to tentatively approach General Relativity). We obtain for $S(a)$ a strictly monotonous function in
the non-negative range of $a$, and $R_{-}> \gamma a_m$ -- specifically $\gamma a_m=0.12058048347$
and $R_{-}=0.12058048511$. As $|\sigma|$ decreases the difference ($R_{-}-\gamma a_m$) increases by the
same order of magnitude.
\par Summarizing, the gravitational collapse of a dust star with mass greater than $M_{\ast}$ produces a
black hole -- defined as the region of spacetime enclosed by the outer event horizon ${R}_{+}$ --
enclosing not a singularity but perpetually bouncing matter in the infinite chain of spacetime maximal extensions inside the outer event horizon.
\par We note that the critical mass depends solely on the parameter $\sigma$.
To have an idea of the order of magnitude of $\sigma$ for event horizon formation, let us
take $M_{\ast} \sim 1.4M_{\bigodot}$, the Chandrasekhar limit. This yields $|\sigma_{\ast}| \sim 10^{-2}~km^{-2}$.
A star with the Chandrasekhar mass would not form an event horizon if the brane tension is smaller than
$10^{-2}~km^{-2}$, so that this value establishes a lower bound for $|\sigma|$. It is also worth
noticing that the geometry (\ref{eq9}) is a solution of brane corrected Einstein's equations $R_{ab}+E_{ab}=0$,
where $E_{ab}$ is the traceless projection of the Weyl curvature on the brane\cite{maeda}; we can easily
extend Birkhoff theorem for these equations.
\par
{\it{Observational Consequences and Final Comments}} -
For the black hole geometry (\ref{eq9}) we have derived the
two classical tests of General Relativity, the planetary perihelia precession and the bending of light.
In a perturbative procedure\cite{anderson} we obtain the perihelion advance per revolution
\begin{eqnarray}
\nonumber
\Omega=\frac{6 \pi G_{N}^{2} M^2}{h^2}\Big( 1-\frac{3 G_{N} M^2}{4 \pi |\sigma| h^4}
- \frac{15 G_{N}^{3} M^4}{8 \pi |\sigma| h^6}\Big),
\end{eqnarray}
with $h^2=G_{N} M a (1-e^2)$, where $e$ is the eccentricity of the orbit and $a$ its semimajor axis.
The first term on the RHS corresponds to the General Relativity correction. Analogously
for the bending of light rays passing in the neighborhood of a spherical massive body, the deflection
angle of the asymptotes is given by
\begin{eqnarray}
\nonumber
\Delta= \frac{4 G_{N} M}{R}\Big(1- \frac{45}{256}~\frac{M}{|\sigma| R^3} \Big)
\end{eqnarray}
where $R$ is the radius of the body. Again the first term on the RHS corresponds to the
General Relativity correction.
For $M=M_{\bigodot}$ the corrections to the effects predicted by General Relativity are neglegible for $|\sigma|$ above the lower limits previously evaluated. However these corrections might be important for masses and/or scales much larger than that of the solar system.
\par In the case of Hawking radiation\cite{hawking}, the
calculations give qualitatively analogous results, with a Planckian thermal spectrum of created particles,
but with a corrected Hawking temperature depending on $M$ and $|\sigma|$,
\begin{eqnarray}
\label{eq12}
T_H=\frac{\Big(R_{+}-R_{-}\Big)}{2 \pi R_{+}~\zeta} ,
\end{eqnarray}
where
\begin{eqnarray}
\nonumber
\zeta&=&\Big[\frac{R_{+}^{3}}{(R_{+}^{2}+ A R_{+}+B)}+\\
\nonumber
&&\Big(\frac{G_{N} M ~(2R_{+}^{3}-3 M/4\pi|\sigma|)(R_{+}-R_{-})^2}{4 (R_{+}-3G_{N} M/2)}
\Big)^{1/2}\Big].
\end{eqnarray}
We note that in the extremal case we have $R_{+}=R_{-}=3 G_{N} M/2$ implying that
$T_{H} \rightarrow 0$ continuously as $R_{+} \rightarrow R_{-}$. The constants $A$ and $B$
have the expression $A=(R_{+}+R_{-})-2G_N M$ and $B=3 G_N M^2/(4 \pi |\sigma|R_{+}R_{-})$.
The observation of Hawking radiation could, in principle, allows to test our results for finite $|\sigma|$.
\par
{\it{Black Hole Thermodynamics}} -
In this section, some thermodynamical results are derived for the case of quasi-extremal black holes. In this case
the roots of our polynomial $P(R)$ are given by $R_{\pm}\simeq 3G_{N}M/2 \pm \epsilon$, where
$\epsilon$ is a small deviation from the extremal case and given by
\begin{eqnarray}
\nonumber
\epsilon=\Big( \frac{3}{8} G_{N}^{2}M^2-\frac{1}{6 \pi G_{N}|\sigma|} \Big)^{1/2}.
\end{eqnarray}
By defining the outer horizon area as $A_{\rm outer}:=4\pi R^2_{+}$ we obtain that
\begin{eqnarray}
\label{A}
dA_{\rm outer} \simeq \frac{9\pi G^3_{N}M_{*}^2}{2~\epsilon}~dM+\frac{M_{*}}{|\sigma_{*}|^2~\epsilon}~d|\sigma|,
\end{eqnarray}
calculated about the extremal configuration($M_{*},|\sigma_{*}|$).
In this approximation the Hawking temperature (\ref{eq12}) assumes the form
\begin{eqnarray}
\nonumber
T_H \simeq \frac{8}{9 \pi} {\frac{\epsilon}{G_{N}^{2} M_{*}^{2}}}~,
\end{eqnarray}
the substitution of which in (\ref{A}) results
\begin{eqnarray}
\label{C}
\frac{1}{4G_{N}}dA_{\rm outer}\simeq\frac{1}{T_{H}}\Big(dM+\frac{M_{*}}{2|\sigma_{*}|}d|\sigma|\Big).
\end{eqnarray}
We can therefore associate the horizon area of the quasi-extremal black hole with the entropy
$S=A_{\rm outer}/4G_{N}$, in accordance to Bekenstein's definition\cite{beken}, turning
Equation (\ref{C}) into an extended Second Law of Thermodynamics, with an extra work term connected to the variation of the brane tension. 
\par {\it{Final Comments}} - Finally, according to previous
work on perturbations in bouncing matter\cite{maier1}, nonlinear resonance mechanisms
typically may occur in the collapsed bouncing matter due to its parametric fluctuations
as produced for instance by a perturbative scalar field. Such a mechanism can
turn the black hole configuration unstable leading to a disruptive ejection
of mass. In fact, the dynamics of oscillatory perturbations of the interior spacetime geometry and
matter content present small domains in the parameter space where the dynamics
is resonant. The remaining parametric space corresponds to stable configurations.
The parameters $|\sigma|$ and the amplitude of the perturbations regulate the
driving of the system from one region to the other. We will approach this issue
in a future publication.
\par
Also the presence of Cauchy horizons $R_{-}$ in the maximal analytical extension of the geometry
(\ref{eq9}) poses a question on the stability of the spacetime. Due to its similarity with
the Reissner-Nordstrom spacetime we should expect that the flux of energy of test fields
may diverge on crossing $R_{-}$ (cf. \cite{chandras} and references therein). However a complete
treatment of the problem should include higher-order nonlinear terms to provide sufficient
conditions for instability. This work is in progress and will also be the object of a future publication.
\par
An analogous problem -- the end state of a collapsing homogeneous scalar field -- was attacked
by Bojowald et al. \cite{bojo1} in the realm of loop quantum
cosmology. In their paper the singularity avoidance basically arises from semiclassical corrections
in the geometric density $1/a^3$, which does not diverge as $a \rightarrow 0$.
However the interior spacetime containing the scalar field
distribution cannot be directly matched to a static Schwarzschild exterior,
leaving therefore the dynamics of horizon formation and its structure undetermined.
A critical threshold scale for horizon formation is obtained, indicating the possibility of
formation of dynamical horizons, but the authors were not able to compute the
associated critical mass because the exterior dynamics remains undetermined.
Their claim that both horizons will become timelike and evaporate is not sustained
by the calculations in our model.
Also several proposals for the construction of nonsingular black holes
in General Relativity appeared in the literature. However none of them are solutions of Einstein's
equations neither are they associated with known physical sources (cf. \cite{hayward} and references therein).


\begin{thebibliography}{99}
\bibitem{comment} We remark that the effective 4-dim cosmological constant can be set zero
in the present case of an extra timelike dimension, by properly fixing the bulk cosmological constant.
\bibitem{maier} Specifically $F$ satisfying
%
{\small
\begin{eqnarray}
\nonumber
\partial F/\partial S= \Big(\Big[\frac{a {\dot{a}}}{1-r^2(k+{\dot{a}}^2)} \Big]\Big[ \frac{(1-k r^2)^{1/2}}
{k(S-\delta)}\Big]\Big) {\Big |}_{r=\gamma}.
\end{eqnarray}
}
\noindent For $|\sigma| \rightarrow \infty$ and $k>0$ we may recover $S=a$ at $r=\gamma$, which corresponds to the Oppenheimer-Snyder matching\cite{weinberg}.
\end{thebibliography}
\end{document}